\documentclass[a4paper,11pt]{article}

\usepackage{a4wide}
\usepackage{amsmath}
\usepackage{graphicx}
\usepackage{pictex}

\title{Neutrino current in a gravitational\\plane wave collision background\vspace{5mm}}
\author{
Donato \textsc{Bini}$^{\;\dag\,\S\;}$, Christian
\textsc{Cherubini}$^{\;\ddag\,\S\P\;}$, Gianluca
\textsc{Cruciani}$^{\;\natural\,\S\;}$ and Andrea
\textsc{Lunari}$^{\;\flat\;\S\;}$\\[5mm]\small
$^\dag$\hspace{2mm}\textit{Istituto per le Applicazioni del Calcolo "Mario Picone", C.N.R.,
I--00161 Rome, Italy}\\[1mm]\small
$^\ddag$\hspace{2mm}\textit{Department of Physics "E.R.
Caianiello", University of Salerno, I--84081, Italy}\\[1mm]\small
$^\P$\hspace{2mm}\textit{Institute of Cosmology and Gravitation,
University of Portsmouth, Portsmouth, PO1 2EG, UK}\\[1mm]\small
$^\natural$\hspace{2mm}\textit{Department of Physics, University
of Perugia, I--06100,
Italy}\\[1mm]\small
$^\flat$\hspace{2mm}\textit{Department of Physics, University of Como, I--22100,
Italy}\\[1mm]\small
$^\S$\hspace{2mm}\textit{ICRA - International Centre for
Relativistic Astrophysics, I--00185 Rome, Italy} }\normalsize

\date{December 23, 2003}

\begin{document}

\maketitle

\begin{flushleft}\small
PACS: 0420C\\
Accepted by {\it International Journal of Modern Physics D}\\
on November 19, 2003
\end{flushleft}\normalsize

\begin{abstract}
The behaviour of a massless Dirac field on a general spacetime
background representing two colliding gravitational plane waves is
discussed in the Newman-Penrose formalism. The geometrical
properties of the neutrino current are analysed and explicit
results are given for the special Ferrari-Iba\~nez solution.
\end{abstract}

\section{Introduction}
Exact solutions of the Einstein equations representing colliding
gravitational plane waves have been  discussed extensively in the
literature and the status of research on this topic is still best
summarized in a monograph by Griffith\cite{griff} published in
1991.

The interest in  such solutions is mainly due to the possibility
they offer to  understand better some nonlinear features of the
gravitational interaction. In general, the spacetime geometry
associated with two colliding gravitational plane waves is very
rich: for example either a spacetime singularity\cite{kha-pe} or a
Killing-Cauchy horizon\cite{Fe-Ib,Fe-Ibbis,Fe-Ib2} can  result
from the nonlinear wave interaction. In general, such spacetimes
contain four regions: a Minkowki region, representing the
initially flat situation before the passage of the two oppositely
directed plane waves, two Petrov type N regions, corresponding to
the waves before the interaction, and an interaction region,
generally of Petrov type I. Two commuting spacelike Killing
vectors are always present, associated with the plane symmetry
assumed for the two colliding waves.

Here we consider a massless Dirac field (neutrino) interacting
with the colliding plane waves. Explicit results are given for the
degenerate solutions found by Ferrari and Iba\~
nez\cite{Fe-Ibbis}, in which the interaction region of the
colliding waves propagating along the common $z$ direction is of
Petrov type D, with a Killing-Cauchy horizon formed at a finite
distance from the collision plane, after a finite time from the
instant of collision. For this spacetime a careful analysis of
timelike geodesics has only recently been performed\cite{bcl}.

Some special features of the neutrino current are discussed in
detail, and previous work of Dorca and
Verdaguer\cite{do-ve1,do-ve2} and Yurtsever\cite{yurt} valid for a
scalar field interacting with gravitational waves is extended here
to spin $1/2$ massless particles.

\section{Colliding waves: the Dirac equation}

The most general form of the spacetime metric  representing two
colliding waves with parallel polarization can be written as
\begin{equation} \label{metgen} \quad ds^2=2\,g_{12}\,dx_1\,dx_2-
g_{33}\,(dx_3)^2-g_{44}\,(dx_4)^2\ ,
\end{equation} where $g_{12},\,g_{33},\,g_{44}$ are real functions
of the  null coordinates $(x_1,\,x_2)$. The complete spacetime
description of the collision must patch together two single-wave
regions (the approaching waves) plus a portion of flat spacetime
corresponding to the situation before the passage of the waves and
the nontrivial interaction region where the metric depends on both
$x_1$ and $x_2$. We distinguish the single-wave metrics by their
dependence on a single null coordinate ($g_{ij}=g_{ij}(x_1)$ for
the progressive wave and $g_{ij}=g_{ij}(x_2)$ for the regressive
one), while in flat spacetime these components are constants,
$g_{12}=2,\,g_{33}=g_{44}=1$. Khan and Penrose \cite{kha-pe}
identified a standard procedure to extend continuously the metric
over all the various regions by a proper use of Heaviside step
functions. We will discuss the details of this extended spacetime
in the case of the explicit solution of Ferrari-Iba\~nez in the
next section.

A wide variety of exact solutions can be found in the literature,
the analysis of which is extremely simplified using the
Newman-Penrose formalism~\cite{New-Pen}. Here the following NP
null frame can be introduced \begin{equation}\label{NPframegen}
l=\partial_{x_2}\ ,\quad n=1/g_{12}\,\partial_{x_1}\ ,\quad
m=\frac{1}{\sqrt{2}}\;\left[\frac{1}{\sqrt{g_{33}}}\,\partial_{x_3}-i\,\frac{1}{\sqrt{g_{44}}}\,
\partial_{x_4}\right]\ . \end{equation} It is convenient to introduce the
following two quantities \begin{equation}
\Xi=\sqrt{g_{12}/(2\,g_{33})}\ ,\qquad
\Phi=\sqrt{g_{12}/(2\,g_{44})} \end{equation} which play a role in
a conformally rescaled form of the metric (\ref{metgen})
\begin{equation} \label{metconf}  \quad ds^2=\frac{\,g_{12}}{2}\,\left[4 dx_1\,dx_2-\Xi^{-2}
\,(dx_3)^2-\Phi^{-2} \,(dx_4)^2\right]\ . \end{equation}

The Dirac equation for massless spin $1/2$ particles then reduces
to
\begin{eqnarray}
\label{diraceq}
&&(D+\epsilon-\rho)\,F_1+(\delta^*+\pi-\alpha)\,F_2=0\ , \nonumber \\
&&(\delta+\beta-\tau)\,F_1+(\Delta +\mu-\gamma)\,F_2=0\ ,
\end{eqnarray}
where the notation and conventions for the NP formalism follow
those of Chandrasekhar\cite{chandra}. A much simplified form of
(\ref{diraceq}) results from rescaling the spin wave functions
$F_1$  and $F_2$  as follows
\begin{eqnarray} \label{rescalgen}
 \quad
F_1 & \rightarrow & (g_{33}\,g_{44})^{-1/4}/\sqrt{g_{12}}\;H_1=(-g)^{-1/4}\;H_1\ , \nonumber \\
 \quad F_2 & \rightarrow & (g_{33}\,g_{44})^{-1/4}\;H_2\
.
\end{eqnarray}
According to the Khan-Penrose procedure\cite{kha-pe}, we must
first examine the Dirac equation for each of the four regions.

\subsection{Case 1: $g_{ij}=g_{ij}(x_1,\,x_2)$}\label{c1}
The NP spin coefficients have a very simple and symmetric form in
the general case
\begin{eqnarray}\label{spincoeff}
  \quad \rho =
-\frac14\;\partial_{x_2}\log\,(g_{33}\,g_{44}) & \ ,\quad
& \sigma = -\frac14\;\partial_{x_2}\log\,(\frac{g_{33}}{g_{44}}) \ ,\nonumber \\
  \quad \lambda =
-\frac{1}{4\,g_{12}}\;\partial_{x_1}\log\,(\frac{g_{33}}{g_{44}})
& \ ,\quad & \mu = -\frac{1}{4\,g_{12}}\;\partial_{x_1}\log\,(g_{33}\,g_{44})\ , \nonumber \\
  \qquad\quad\ \epsilon = \frac12\;\partial_{x_2}(\log
g_{12}) & \ ,\quad & \alpha=\beta=\gamma=\kappa=\nu=\pi=\tau=0\ ,
\end{eqnarray}
while the Weyl scalars are
\begin{eqnarray}
\Psi_0=D\,\sigma-2\,\sigma\,(\rho+\epsilon)&=&\frac18\,\bigg[
[\partial_{x_2}\log\,(g_{12}\,g_{33})]^2-[\partial_{x_2}\log\,(g_{12}\,g_{44})]^2 \nonumber \\
&&\left. -2\left(\frac{\partial_{x_2}^2\,g_{33}}{g_{33}}-
\frac{\partial_{x_2}^2\,g_{44}}{g_{44}}\right)\right] ,\nonumber \\
  \Psi_1=-\delta\,\epsilon&=&0\ ,\nonumber \\
\Psi_2=\frac{D\,\mu-\Delta\,\epsilon+2(\mu\,\epsilon-\lambda\,\sigma)}
{3}&=&\frac{1}{12\,g_{12}}
\left(2\,\partial_{x_1}\log\,g_{12}\,\partial_{x_2}
\log\,g_{12} \right. \nonumber \\
&&-\frac12\,\partial_{x_1}\log\,(g_{33}\,g_{44})\,\partial_{x_2}
\log\,(g_{33}\,g_{44}) \nonumber \\
&&\left.
-2\,\frac{\partial_{x_1}\,\partial_{x_2}\,g_{12}}{g_{12}}+
\frac{\partial_{x_1}\, \partial_{x_2}\,g_{33}}{g_{33}}+
\frac{\partial_{x_1}\,\partial_{x_2}\,g_{44}}{g_{44}}\right)\ ,\nonumber \\
\Psi_3&=&0 \ ,\nonumber \\
\Psi_4=-\Delta\,\lambda-2\,\lambda\,\mu&=&\frac{1}{8\,{g_{12}}^2}\,
\bigg[
[\partial_{x_1}\log\,(g_{12}\,g_{33})]^2-[\partial_{x_1}\log
\,(g_{12}\,g_{44})]^2 \nonumber \\
&&\left. -2\left(\frac{\partial_{x_1}^2\,g_{33}}{g_{33}}-
\frac{\partial_{x_1}^2\,g_{44}}{g_{44}}\right)\right]\ .
\end{eqnarray}

Since the metric is independent of $(x_3,\ x_4)$ one can seek
solutions (normal modes) of the form \begin{equation} H_1=e^{i\,
(K_3\,x_3+K_4\,x_4)}\,h_1(x_1,\,x_2)\ ,\ \ H_2=e^{i\,
(K_3\,x_3+K_4\,x_4)}\,h_2(x_1,\,x_2)\ , \end{equation} where $K_3$
and $K_4$ are real constants and $h_1$, $h_2$ will depend on them.
The Dirac equation then becomes
\begin{eqnarray} \label{dirac1gen}
 \quad \frac{\partial h_1(x_1,\,x_2)}{\partial x_2} & = &
\left(K_4\,\Phi -i
\,K_3\, \Xi \right)h_2(x_1,\,x_2)\ , \nonumber \\
 \quad \frac{\partial h_2(x_1,\,x_2)}{\partial x_1} & = &
\left(-K_4\,\Phi-i \,K_3\, \Xi \right)h_1(x_1,\,x_2)\ ,
\end{eqnarray}
and the general solution can be given as superposition of modes
\begin{equation} \mathcal{H}_1=\int dK_3\, dK_4\, H_1\ , \qquad
\mathcal{H}_2=\int dK_3 \, dK_4 \, H_2\ . \end{equation}

\subsection{Case 2: $g_{ij}=g_{ij}(x_1)$}\label{c2}
In this case, only $\lambda$ and $\mu$ remain nonzero among the
spin coefficients (\ref{spincoeff}). Since the metric is
independent of $(x_2,\ x_3,\ x_4)$ one can seek solutions of the
form \begin{equation} H_1=e^{i\,(K_2\, x_2 +K_3\, x_3+K_4\,
x_4)}\,h_1(x_1)\ ,\ \ H_2=e^{i\,(K_2\, x_2 +K_3\, x_3+K_4\,
x_4)}\,h_2(x_1)\ , \end{equation} where $K_2$, $K_3$ and $K_4$ are
real constants and $h_1$ and $h_2$ will also depend on them. The
Dirac equation then becomes
\begin{eqnarray} \label{dirac2gen}
 \quad
h_1(x_1) & = & \frac{1}{K_2}\,\left(-K_3\, \Xi -i\,K_4\, \Phi \right)h_2(x_1)\ , \nonumber \\
 \quad h_2'(x_1) & = &
\frac{i}{K_2}\left[(K_3\,\Xi)^2+(K_4\,\Phi)^2\right]h_2(x_1)\ ,
\end{eqnarray}
and the general solution can be given as superposition of these
modes \begin{equation} \mathcal{H}_1=\int dK_2\, dK_3\, dK_4\,
H_1\ ,\qquad \mathcal{H}_2=\int dK_2\, dK_3 \, dK_4 \, H_2\ .
\end{equation}

\subsection{Case 3: $g_{ij}=g_{ij}(x_2)$}\label{c3}
Here the non-zero spin coefficients are $\rho$, $\sigma$ and
$\epsilon$. Since the metric is independent of $(x_1,\ x_3,\ x_4)$
one can seek solutions of the form \begin{equation}
H_1=e^{i\,(K_1\, x_1 +K_3\, x_3+K_4\, x_4)}\,h_1(x_2)\ ,\ \
H_2=e^{i\,(K_1\, x_1 +K_3\, x_3+K_4\, x_4)}\,h_2(x_2)\ ,
\end{equation} where $K_1$, $K_3$ and $K_4$ are real constants and
as in the previous cases $h_1$ and $h_2$ will also depend on them.
The Dirac equation then becomes:
\begin{eqnarray} \label{dirac3gen}
 \quad
h_1'(x_2) & = & \frac{i}{K_1}\left[(K_3\,\Xi)^2+(K_4\,\Phi)^2\right]h_1(x_2)\ , \nonumber \\
 \quad h_2(x_2) & = & \frac{1}{K_1}\,\left(-K_3\,\Xi
+i\,K_4\,\Phi \right)h_1(x_2)\ ,
\end{eqnarray}
and the general solution can be given as superposition of these
modes: \begin{equation} \mathcal{H}_1=\int dK_1\, dK_3\, dK_4\,
H_1\ , \qquad \mathcal{H}_2=\int dK_1\, dK_3 \, dK_4 \, H_2\ .
\end{equation}

\subsection{Case 4: $g_{ij}=$ (positive) constants}\label{c4}
For a metric with constant components the spin coefficients are
obviously all zero. We assume without any loss of generality:
$g_{12}=2$ and $g_{33}=g_{44}=1$. Since the metric is constant one
can seek solutions of the form \begin{equation} H_1=e^{i\,(K_1\,
x_1 +K_2\, x_2+K_3\, x_3+K_4\, x_4)}\,h_1\ ,\ \ H_2=e^{i\,(K_1\,
x_1 +K_2\, x_2+K_3\, x_3+K_4\, x_4)}\,h_2\ , \end{equation} with
$h_1$ and $h_2$ also constants. The Dirac equation reduces to
\begin{eqnarray} \label{dirac4gen}
 \quad
K_2\,h_1+(K_3+i\,K_4)\,h_2 & = & 0\ , \nonumber \\
 \quad (K_3-i\,K_4)\,h_1+K_1\,h_2 & = & 0\ ,
\end{eqnarray}
which implies (by setting to zero the determinant of the
coefficient matrix of the linear system (\ref{dirac4gen}))
\begin{equation}\label{coeffdet}  \quad K_1\,K_2=K_3^{\;2}+K_4^{\;2}\ . \end{equation}

In the next section we will specialize these cases to the
horizon-forming Ferrari-Iba\~nez metric and discuss their
solutions.

\section{The horizon-forming Ferrari-Iba\~nez metric}

In 1987 Ferrari and Iba\~nez\cite{Fe-Ibbis,Fe-Ib2} found a type D
solution of the Einstein equations that can be interpreted as
describing the collision of two linearly polarized gravitational
plane waves propagating along a common direction $z$ in opposite
senses and developing a non-singular Killing-Cauchy horizon when
they collide. Using the standard  coordinates $(t,\,z,\,x,\,y)$,
with an appropriate choice of the amplitude parameters associated
with the strength of the waves, the metric takes the form:
\begin{equation}  \qquad \label{met1} ds^2=(1+ \sin
t)^2\,(dt^2-dz^2)-\frac{1-\sin t}{1+\sin t}\;dx^2- \cos^2
z\,(1+\sin t)^2\,dy^2\ . \end{equation} The interaction region
where this form of the metric is valid (designated as \lq\lq
Region I", as in most of the literature\cite{do-ve1}) is
represented in the $(t,\, z)$ diagram by a triangle whose vertex
(representing the initial event of collision) can be identified
with the origin of the coordinate system; the horizon is mapped
onto the base of the shaded triangle in Fig.~1.

In order to describe the larger spacetime of which this is only
one region, one must introduce the two null coordinates
\begin{equation} u=(t-z)/2 , \ v=(t+z)/2 \Longleftrightarrow\
t=u+v ,\ z=v-u\ , \end{equation} in terms of which the metric
(\ref{met1}) takes the form
\begin{eqnarray}\label{met2}
 \qquad
ds^2&=&4\;[1+\sin(u+v)]^2 du\,dv \nonumber \\
 \qquad
&&-\frac{1-\sin(u+v)}{1+\sin(u+v)}\;dx^2-\cos^2(u-v)
[1+\sin(u+v)]^2dy^2 \ .
\end{eqnarray}

Following Khan-Penrose\cite{kha-pe} one can easily extend the
formula for the metric from the interaction region to the
remaining parts of the spacetime representing the single wave
zones and the flat spacetime zone before the waves arrive. The
interaction region corresponds to the triangular region in the
$(u,\,v)$ plane bounded by the lines $u=0$, $v=0$ and $u+v=\pi/2$.
One need only make the following substitutions in (\ref{met2}):
\begin{equation} \label{k-p} u\,\rightarrow\,u\;H(u)\qquad\qquad
v\,\rightarrow\,v\;H(v)\nonumber \end{equation} which give rise to
the four regions \begin{equation}
\begin{array}{lll} \quad\quad
 u\geq 0\,,\;v\geq 0\,, u+v<\pi/2 \qquad& \mbox{Region I} \qquad& \mbox{Interaction region}\\
  \quad\quad 0\leq u < \pi/2 \,,\;v<0 & \mbox{Region II} & \mbox{Single $u$-wave region}\\
  \quad\quad u<0\,,\;0\leq v < \pi/2 & \mbox{Region III} & \mbox{Single $v$-wave
 region}\\
  \quad\quad u<0\,,\;v<0 & \mbox{Region IV} & \mbox{Flat space}
\end{array}
\end{equation} shown in fig. 1. In this way the extended metric in general
is $C^0$ (but not $C^1$) along the null boundaries $u=0$ and
$v=0$. It is worth noting that certain calculations are more
easily done in one or the other of these two sets of coordinates,
so we will switch back and forth between them as needed.

\begin{figure}[t]
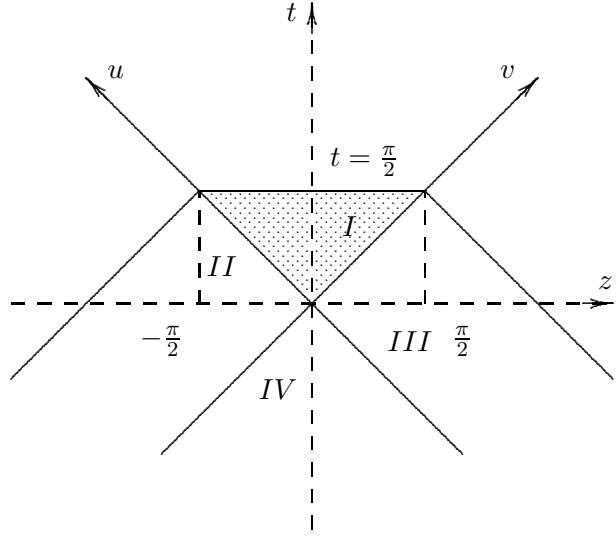

$$
\typeout{pictex fig1}
 \beginpicture
  \setcoordinatesystem units <1cm,1cm> point at 0 0
  \setdashes
   \putrule from  -4   0      to    4  0
   \putrule from   0  -3      to    0  4

\putrule from 1.5 0 to 1.5 1.5 \putrule from -1.5 0 to -1.5 1.5
  \setsolid \setlinear
    \putrule from -1.5 1.5 to 1.5 1.5
    \plot  0  0    3 3    /
    \plot  0  0   -3 3    /
    \plot -2 -2    0 0    /
    \plot  2 -2    0 0    /
    \plot  4 -1  1.5 1.5  /
    \plot -4 -1 -1.5 1.5  /
  \setshadegrid span <.025in>
   \vshade -1.5 1.5 1.5   0 0 1.5   1.5 1.5 1.5 /
  \put {$z$}               [rb]     at      4  0.2
  \put {$t$}               [rt]     at   -0.2  4
  \put {$u$}               [rrb]    at   -2.5  3
  \put {$v$}               [rlb]    at    2.5  3
  \put {$I$}               [rb]     at    0.6  0.9
  \put {$II$}              [rr]     at     -1  0.5
  \put {$III$}             [rl]     at      1 -0.5
  \put {$IV$}              [rt]     at   -0.2 -1

 \put {$t=\frac{\pi}{2}$}             at    0.7   1.9
 \put {$\frac{\pi}{2}$}               at      2  -0.5
 \put {$-\frac{\pi}{2}$}              at     -2  -0.5

  \arrow <.3cm> [.1,.4]    from    3.6    0  to   4 0
  \arrow <.3cm> [.1,.4]    from      0  3.6  to   0 4
  \arrow <.3cm> [.1,.4]    from   2.72 2.72  to   3 3
  \arrow <.3cm> [.1,.4]    from  -2.72 2.72  to  -3 3
\endpicture
$$
\caption{\it The null coordinates and the different regions they
induce.}\label{fig:1}
\end{figure}

We can apply the results of the previous section using the
coordinates  $(u,\,v,\,x,\,y)$ and the extended metric
(\ref{met2}) -- (\ref{k-p}).

\subsection{Region I: wave interaction}

The metric
\begin{eqnarray}\label{reg1}
 \quad
g_{12} &=& 2\,[1+\sin(u+v)]^2\ ,\nonumber \\
 \quad
g_{33} &=& \frac{1-\sin(u+v)}{1+\sin(u+v)}\ ,\nonumber \\
 \quad g_{44} &=& \cos^2(v-u)\,[1+\sin(u+v)]^2\ .
\end{eqnarray}
is of Petrov type D, the NP frame (\ref{NPframegen}) is not
principal (i.e. not aligned along the two repeated principal null
directions of the spacetime) and the Dirac equation
(\ref{dirac1gen}) becomes
\begin{eqnarray}\label{diracreg1}
 \quad \frac{\partial h_1(u,\,v)}{\partial v} &=&
\left(\frac{K_y}{\cos(v-u)}
-i\,K_x\frac{[1+\sin(u+v)]^2}{\cos(u+v)}\right)\,h_2(u,\,v)\ , \nonumber \\
 \quad \frac{\partial h_2(u,\,v)}{\partial u} &=&
\left(-\frac{K_y}{\cos(v-u)}
-i\,K_x\frac{[1+\sin(u+v)]^2}{\cos(u+v)}\right)\,h_1(u,\,v)\ ,
\end{eqnarray}
where a convenient notational change has been made with respect to
the previous section, denoting $K_{1,2,3,4}$ as $K_{u,v,x,y}$.

To exploit the symmetries of  (\ref{diracreg1}) we again transform
the Dirac variables
\begin{eqnarray}
h_1(u,\,v) & = & s_1(u,\,v)+s_2(u,\,v)\ , \nonumber \\
h_2(u,\,v) & = & s_1(u,\,v)-s_2(u,\,v)\ ,
\end{eqnarray}
and then we switch back to the spacetime coordinates
$(t,\,z,\,x,\,y)$, introducing the abbreviated notation
\begin{displaymath}
\alpha(t)=\frac{(1+\sin t)^2}{\cos t}\ , \quad
\beta(z)=\frac{1}{\cos z}\ .
\end{displaymath}
Equations (\ref{diracreg1}) then become
\begin{eqnarray}\label{dirac-int-2}
 \quad \frac{\partial s_1(t,\,z)}{\partial
t}+\frac{\partial s_2(t,\,z)}{\partial z}
& = & -i\,K_x\,\alpha(t)\,s_1(t,\,z)-K_y\,\beta(z)\,s_2(t,\,z)\ , \nonumber \\
 \quad \frac{\partial s_1(t,\,z)}{\partial
z}+\frac{\partial s_2(t,\,z)}{\partial t} & = &
K_y\,\beta(z)\,s_1(t,\,z)+i\,K_x\,\alpha(t)\,s_2(t,\,z)\ .
\end{eqnarray}
These equations can be separated by assuming
$s_1(t,\,z)=T_1(t)Z_1(z)$ and $s_2(t,\,z)=T_2(t)Z_2(z)$. In fact
dividing the first by $T_2(t)\,Z_1(z)$ and the second by
$T_1(t)\,Z_2(z)$, we obtain
\begin{eqnarray}
\frac{T_1'(t)}{T_2(t)}+\frac{Z_2'(z)}{Z_1(z)}+i\,K_x\,\alpha(t)\,
\frac{T_1(t)}{T_2(t)}+K_y\,\beta(z)\frac{Z_2(z)}{Z_1(z)} & = & 0\ , \nonumber \\
\frac{T_2'(t)}{T_1(t)}+\frac{Z_1'(z)}{Z_2(z)}-i\,K_x\,\alpha(t)\,
\frac{T_2(t)}{T_1(t)}-K_y\,\beta(z)\frac{Z_1(z)}{Z_2(z)} & = & 0\
,
\end{eqnarray}
and hence
\begin{eqnarray}
\label{TZ}
T_1'(t)+i\,K_x\,\alpha(t)\,T_1(t) & = & -K\,T_2(t)\ , \nonumber \\
T_2'(t)-i\,K_x\,\alpha(t)\,T_2(t) & = & -J\,T_1(t)\ , \nonumber \\
Z_1'(z)-K_y\,\beta(z)\,Z_1(z) & = & J\,Z_2(z)\ , \nonumber \\
Z_2'(z)+K_y\,\beta(z)\,Z_2(z) & = & K\,Z_1(z)\ ,
\end{eqnarray}
where J and K are arbitrary constants. By differentiating these
equations the functions $T, Z$ are seen to satisfy a system of
four completely separated second-order differential equations of
the following form: \begin{equation}\label{secord}
\frac{df(x)}{dx^2}-[a^2\,\mathcal{A}(x)^2+a\,\mathcal{A}'(x)+J\,K]\,f(x)=0\
,
\end{equation} with the respective sets of values for
$f(x),\,a,\,\mathcal{A}\,$: \begin{equation} \label{faA}
f(x)\Rightarrow\left\{\begin{array}{l}T_1\\T_2\\Z_1\\Z_2\end{array}\right.\qquad
a\Rightarrow\left\{\begin{array}{r}-i\,K_x\\i\,K_x\\K_y\\-K_y\end{array}\right.\qquad
\mathcal{A}\Rightarrow\left\{\begin{array}{l}\alpha(t)\\\alpha(t)\\\beta(z)\\\beta(z)\end{array}
\ . \right. \end{equation} With the restriction
$\{J\,K=-\left(\frac{n}{2}\right)^2,\,n\in N\}$, the
(\ref{secord}) for $Z_1,\,Z_2$ can be integrated analytically. In
fact with the variable change $z=\pi/2-\theta$ and the rescaling
$Z_{1,\,2}(\theta)=\sqrt{\sin\theta}\;\Theta_{1,\,2}(\theta)$ the
resulting equation for $\Theta_{1,\,2} (\theta)$ becomes
\begin{equation}\label{teukst}  \quad
\Theta''(\theta)+\cot\theta\,\Theta'(\theta)+\left[\frac{2\,s\,m
\cos\theta-m^2-s^2}{\sin^2
\theta}+l\,(l+1)\right]\Theta(\theta)=0\ , \end{equation} where
$J\,K = -[l\,(l+1)+s^2]$ and $s=\mp 1/2$ must hold in order to
have regular solutions, with $\Theta_{1}(\theta)$ corresponding to
$s=-1/2$, $m=K_y$ (or $s=1/2$, $m=-K_y$) and $\Theta_{2}(\theta)$
corresponding to $s=1/2$, $m=K_y$ (or $s=-1/2$, $m=-K_y$). A class
of solutions of  (\ref{teukst}) is represented by the {\it
spin-weighted spherical harmonics} ${}_{s}Y^{\pm
K_y}_l(\theta,\,0)$ (first introduced by Newman and
Penrose\cite{New-Pen} as a natural extension of the scalar
harmonics to the Hilbert space of the square-integrable spin-$s$
functions defined on the 2-dimensional unit sphere). An explicit
definition of these functions, together with a list of their main
properties and features  is given in the Appendix. The
corresponding solutions of  Eqs.~(\ref{TZ}--\ref{faA}) can then be
denoted by
\begin{eqnarray}
T_{1}(t)&=& T{}^{K_x}_l (u+v),\nonumber \\
T_{2}(t)&=& T{}^{-K_x}_l (u+v) \nonumber \\
Z_{1}(z)&=& \sqrt{\cos(v-u)}\,{}_{-1/2}Y^{K_y}_l(\pi/2+u-v,\,0),\nonumber \\
Z_{2}(z)&=& \sqrt{\cos(v-u)}\,{}_{1/2}Y^{K_y}_l(\pi/2+u-v,\,0)
\end{eqnarray}
where indices $1,2$ correspond to the signs $\pm$ respectively.
The general solution will then be a superposition of such modes
\begin{eqnarray}
\label{solfinfin}   h_1(u,v)&=& \sqrt{\cos(v-u)}\,
\sum_{l=|K_y|}^\infty C_l
[T{}^{K_x}_l (u+v)\, {}_{1/2}Y^{-K_y}_l(\pi/2+u-v,\,0)+\nonumber \\
  && \qquad \qquad \qquad \qquad T{}^{-K_x}_l (u+v)\,
{}_{1/2}Y^{K_y}_l(\pi/2+u-v,\,0) ]
\end{eqnarray}
with  an analogous expression for $h_2$, where but the functions
$T{}^{\pm K_x}_l (u+v)$ can only be given numerically.

\subsubsection{Teukolsky Master Equation}

We notice here that as this portion of spacetime is Petrov type D
actually one can write, following notation and conventions of
Teukolski\cite{teuk}, a Master Equation for any spin $s=0,\pm 1/2,
\pm 1,\pm 3/2,\pm 2$ perturbations of the form \begin{equation}
\psi(t,z,x,y)=e^{iK_x x}\, e^{iK_y y}\,  T(t)\, Z(z)\
\end{equation} of this background. To do this it is convenient to
adopt a (Kinnersley) principal null tetrad, different from the one
introduced above for the Dirac field, aligned with the two
repeated principal null directions of the spacetime
\begin{eqnarray}
l &=& \frac{1}{2\cos t}\left[ \partial_t +\frac{(1+\sin t)^2}{\cos
t}\,
\partial_x \right]\ ,\nonumber \\
n &=& \frac{\cos t}{(1+\sin t)^2}\, \partial_t -\partial_x \ , \nonumber \\
m &=& \frac{1}{\sqrt{2}\, (1+ \sin t)} \left[\partial_z +
\frac{i}{\cos z}\partial_y\right]\ ,
\end{eqnarray}
so that one gets two decoupled equations, one for  $Z(z)$
\begin{equation} \label{zmotion} \frac{d^2 Z(z)}{dz^2}-
\tan z \frac{dZ(z)}{dz}-V_z(z)Z(z)=0\ , \end{equation} where
\begin{equation} V_z(z)=\frac{K_y^2 - 2 K_y s \sin z + s^2}{\cos^2
z}+\Omega\ , \end{equation} with $\Omega$ a separation constant,
and one for $T(t)$ \begin{equation} \label{eqT} \frac{d^2
T(t)}{dt^2}- (1+2 s)\tan t \frac{dT(t)}{dt} -V_t(t) T(t)=0\ ,
\end{equation} where \begin{equation}  \quad
V_t(t)=\Omega+s(s+1) -K_x^2\frac{(1+\sin t)^4}{\cos^2
t}-2\,iK_x\,s
 \left(  \sin t  +2 \frac{1+ \sin t}{\cos^2 t} \right)\ .
\end{equation} Again,  by introducing the new variable $\theta = \pi/2 +z$
Eq. (\ref{zmotion}) takes the same form of Eq. (\ref{teukst}), and
with  $\Omega=-l(l+1)$, with $l$, $m=K_y$ integers or
half-integers and $s=0,\pm 1/2, \pm 1,\pm 3/2,\pm 2$, its
solutions are the  spin-weighted spherical harmonics, with
spin-weight $s$, generalizing the results already given for the
Dirac field.

To accomplish the task of extending this description all over the
spacetime it is obviously necessary to match the equations for the
various spin fields also in the other two regions of type N and in
the Minkowski portion. This can be done for spin $s=1$ in complete
analogy to what we have done for the $s=1/2$ case. The
gravitational case $s=2$ is to be more carefully handled because
of the nonvanishing radiative contribution of the background
gravitational field in the two N regions. However this analysis
goes beyond the aims of the present paper and it will be studied
subsequently.

\subsection{Region II: single $u$-wave}

The metric
\begin{eqnarray}\label{reg2}
 \quad
g_{12} &=& 2\,(1+\sin u)^2\ ,\nonumber \\
 \quad
g_{33} &=& \frac{1-\sin u}{1+\sin u}\ ,\nonumber \\
 \quad g_{44} &=& \cos^2(u)\,(1+\sin u)^2\ ,
\end{eqnarray}
is of Petrov type N, the NP frame (\ref{NPframegen}) is principal
and the Dirac equation (\ref{dirac2gen}) gives
\begin{eqnarray}\label{diracreg2}
h_1(u) & = & \frac{-K_x\,(1+\sin u)^2-i\,K_y}{K_v\,\cos u}\;\; h_2(u)\ , \nonumber \\
h_2'(u) & = & i\,\frac{K_y^2+K_x^2\,(1+\sin u)^4}{K_v\,\cos^2
u}\;\; h_2(u)\ .
\end{eqnarray}

The ordinary differential equation for $h_2(u)$ can be solved
exactly \begin{equation} \label{sol}
h_2(u)=-\sqrt{2}\,N\,\frac{K_v}{K_x+i\,K_y}\,e^{\frac{i}{K_v}\,
\left[K_x^{2}\,A(u)+K_y^2\,\tan u\right]}\ , \end{equation} where
$N$ is a normalization constant and the function $A(x)$ is defined
by
\begin{equation} \label{Afunct} A(x)=\frac{(\sin x+8)\,(\cos^2
x+2)}{2\,\cos x}+7\,\tan x-\frac{15}{2}\,x-12\ . \end{equation}

\subsection{Region III: single $v$-wave}

The metric
\begin{eqnarray}\label{reg3}
 \quad
g_{12} &=& 2\,(1+\sin v)^2\ ,\nonumber \\
 \quad
g_{33} &=& \frac{1-\sin v}{1+\sin v}\ ,\nonumber \\
 \quad g_{44} &=& \cos^2(v)\,(1+\sin v)^2\
\end{eqnarray}
is again of Petrov type N, the NP frame (\ref{NPframegen}) is
principal and the Dirac equation (\ref{dirac3gen}) is:

\begin{eqnarray}\label{diracreg3}
h_1'(v) & = & i\,\frac{K_x^2\,(1+\sin v)^4+K_y^2}{K_u\,\cos^2 v}\;\; h_1(v)\ , \nonumber \\
h_2(v) & = & -\frac{K_x\,(1+\sin v)^2-i\,K_y}{K_u\,\cos v}\;\;
h_1(v)\ .
\end{eqnarray}
The solution for $h_1(v)$ is \begin{equation}
h_1(v)=\sqrt{2}\,N\,e^{\frac{i}{K_u}\,\left[K_x^{2}\,A(v)+K_y^2\,\tan
v\right]}\ , \end{equation} where $N$ is a normalization constant
and $A(v)$ is defined in (\ref{Afunct}). Here one has the same
problems matching the solution to region I as for region II.

\subsection{Region IV: flat spacetime}

The metric \begin{equation}\label{reg4} g_{12}=2,\qquad g_{33}=
g_{44}= 1\ . \end{equation} is the Minkowski metric. The relation
between the K-constants that comes from the Dirac equation
(\ref{coeffdet}) is \begin{equation} K_u\,K_v = K_x^2 + K_y^2
\end{equation}
(which corresponds to energy-momentum conservation for the test particle\cite{bcl}).\\
A straightforward solution for the amplitudes of the wave
functions is: \begin{equation}  \quad h_1=\sqrt{2}\, N\ ,\qquad
h_2=-\sqrt{2}\,N\,\frac{K_v}{K_x + i\,K_y}=-\sqrt{2}\,N\,\frac{K_x
-i\,K_y}{K_u}\ . \end{equation}

This list of the various cases concludes the analysis of the Dirac
equation. We have seen that an explicit solution can be given
completely in the two type N regions and of course in the flat
Minkowski region. In the type D region the Dirac equation can be
separated in its $(t\, ,z)$ dependence, but only the separated $z$
equation can be solved exactly in terms of spin-weighted spherical
harmonics.

\subsection{Tracking the Dirac field}

It is now clear how to obtain the formal solution of the Dirac
equation in the four spacetime regions. However, since the general
solution in region I can only be given numerically, the problem of
(continuously) matching at the boundaries $I-II$ and $I-III$ can
only be approached numerically. To be more explicit, consider the
example of a transition from region I to region II for $h_1$.
Approaching the $u$ axis ($v\to 0$) one has
\begin{eqnarray}
  \lim_{v\to 0}\, h_1(u,v)&=& \sqrt{\cos u}\,
\sum_{l=|K_y|}^\infty C_l
[T{}^{K_x}_l (u)\, {}_{1/2}Y^{-K_y}_l(\pi/2+u,\,0)+\nonumber \\
  && \qquad \qquad \qquad \qquad T{}^{-K_x}_l (u)\,
{}_{1/2}Y^{K_y}_l(\pi/2+u,\,0) ]\ ,
\end{eqnarray}
and this limit must coincide with the solution $h_1(u)$ of
Eq.~(\ref{diracreg2}) in region II, so that it can be seen as a
series representation for $h_1(u)$ in terms of spin-weighted
spherical harmonics. But the coefficients of this expansion are
not constant and thus we can no longer use the orthonormality
properties of the harmonics to extract the coefficients $C_l$
analytically.

Special cases can be studied numerically. For example, once the
$(K_u, K_v, K_x, K_y)$ parameters are assigned in region IV, the
solution is continuously propagated into regions II and III
without any arbitrariness, reaching the boundaries of region I.

At each crossing point a single mode solution of region I can be
numerically constructed and made continuous with the cited single
wave solutions of region II and III. Even if they could be
superimposed, these single mode solutions can account for a
qualitative description of the behaviour of the Dirac field as
scattered by the gravitational waves.

In the next section the  neutrino current will be analyzed
separately in the four regions of the spacetime following this
procedure.

\section{Neutrino current}

In order to study the properties of the solution and its matching
across the boundary of the various regions of the spacetime, we
analyze the associated neutrino current.

From the Newman-Penrose formalism, the expression for the neutrino
current null vector is the following: \begin{equation}
\label{dircurr} J=2\,[l\,|F_1|^2+n\,|F_2|^2+m\,(F_1\,F_2^*)+\bar
m\,(F_1^*\,F_2)]\ . \end{equation} In terms of the formalism
developed so far the neutrino current density  can be expressed in
the form
\begin{eqnarray}\label{denscurrgen}
 \quad j=\sqrt{-g}\;J&=&
j^u\, \partial_u + j^v\, \partial_v +j^x\, \partial_x +j^y\, \partial_y\, , \nonumber \\
&=& 2\,\{|h_2|^2\,\partial_u+|h_1|^2\,\partial_v+2\,\Xi\, \Re
e(h_1\;h_2^*)\,\partial_x+2\,\Phi\,\Im
m(h_1\;h_2^*)\,\partial_y\}\ ,
\end{eqnarray}
that can easily be specialized to the various regions by
identifying the correct dependence of $h_1,\,h_2,\,\Xi,\,\Phi$.
The current conservation law in the wave interaction region
results in the simple condition
\begin{equation}\label{currentcons}
\partial_u  \, j^u + \partial_v \, j^v =0\, ,
\end{equation} which trivially becomes $\partial_u \, j^u =0$ in region II
and $\partial_v \,  j^v =0$ in region III. Introducing the
(closed) \lq\lq projected current $1-$form\rq\rq \begin{equation}
I=j^u dv - j^v du\, , \qquad dI =0\ \end{equation} allows the use
of Stokes' theorem over a bounded portion $\Omega$ of spacetime
inside region I: $ \int_{\partial\, \Omega} I =0 \ $. In order to
get information on the interaction of the neutrino current with
the colliding gravitational waves we  examine the behaviour of the
neutrino current vector limiting ourselves to the $(u,\,v)$ plane,
drawing the corresponding field lines
\begin{equation} \frac{du}{dv}=\frac{J^u}{J^v}=\frac{j^u}{j^v}=\frac{|h_2|^2}{|h_1|^2}\ .
\end{equation} Along such lines  the $1$-form $I$ vanishes
identically by definition.

This equation can be exactly integrated in the single wave
regions, giving

\begin{equation}   \quad v=\frac{K_y^2}{K_v^2}\, \tan v
+\frac{K_x^2}{K_v^2}\,\hat A (v)+C_1 \end{equation} in  region II
and

\begin{equation}   \quad u=\frac{K_y^2}{K_u^2}\, \tan v
+\frac{K_x^2}{K_u^2}\,\hat A (u)+C_2 \end{equation} in  region
III, where $C_1$ and $C_2$ are integration constants and
\begin{equation} \hat A(x)=A(x)+12 . \end{equation} Moreover, one can show
by explicit calculation that the following remarkable property
holds (in the single wave regions only): the (null) neutrino
current rescaled by the wavefront volume element \begin{equation}
{J_r}^\alpha\,=\,\sqrt{g_{33}\;g_{44}}\,J^\alpha \end{equation} is
geodesic, i.e. satisfies a speciality property as discussed by
Wainwright\cite{wain}.

The lines of the neutrino current are plotted in Figs.~2-4 for
selected values of the (global) parameters $K_x$ and $K_y$ for a
particle entering the single $u$-wave region from the flat
spacetime regime, with different values of  energy (i.e. for
different values of $K_v$). In regions II and III the integral
curves of the current density are null geodesics.

\begin{figure} [b]
\includegraphics[width=150mm,height=75mm]{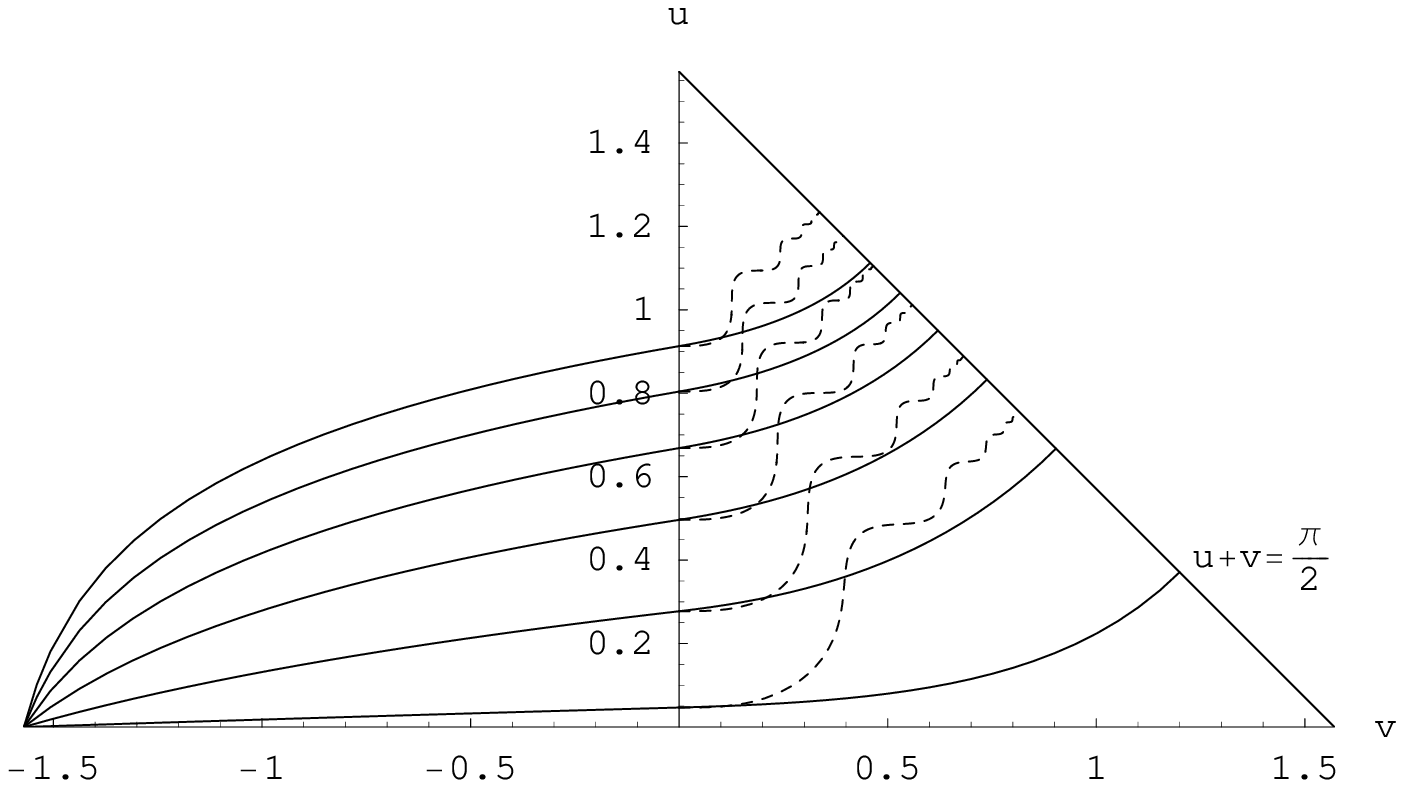}
\vspace{5mm} \caption{Several lines of the current field are
plotted here, corresponding to the case $K_x=1$, $K_y=1/2$,
$0.2<|K_v|<2.2$, $l=1/2$, where the particle is conventionally set
to enter the single-wave zone at the point $(-\pi/2,\,0)$. For
$v<0$ the lines follow geodesic paths in this $(u,\,v)$ slice,
while in region I they decouple (current lines are dashed,
geodesic lines are solid). The current clearly oscillates around
$u-v=const.$ axes, damping its oscillations as the horizon (the
oblique line at the right) is approached.} \label{fig:2}
\end{figure}

\begin{figure} [t]
\includegraphics[width=150mm,height=75mm]{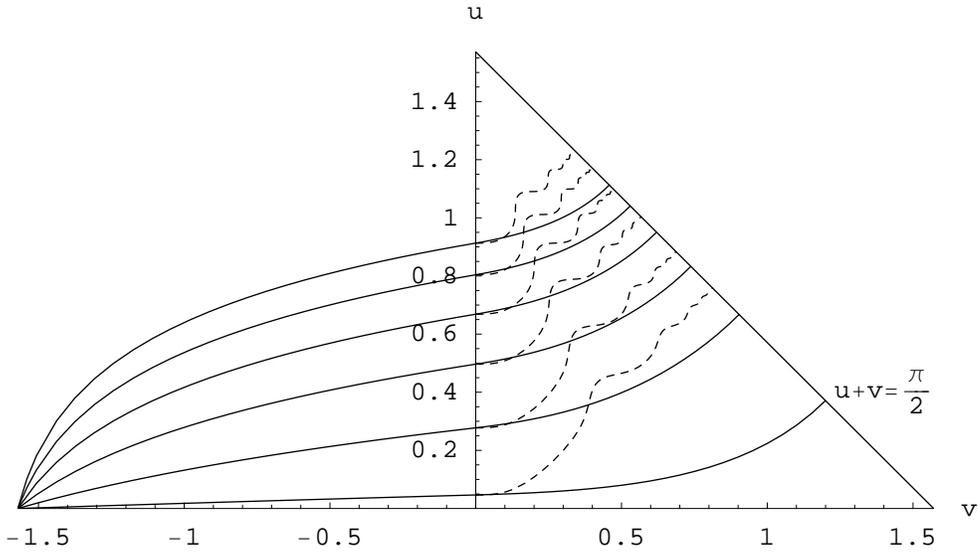}
\vspace{5mm} \caption{Figure 2
repeated for $l=3/2$, showing the influence of this parameter: it
just slightly changes the amplitude of the oscillations, while
having almost no effect on the horizon-crossing value of the field
lines.} \label{fig:3}
\end{figure}

\begin{figure} [t]
\includegraphics[width=150mm,height=75mm]{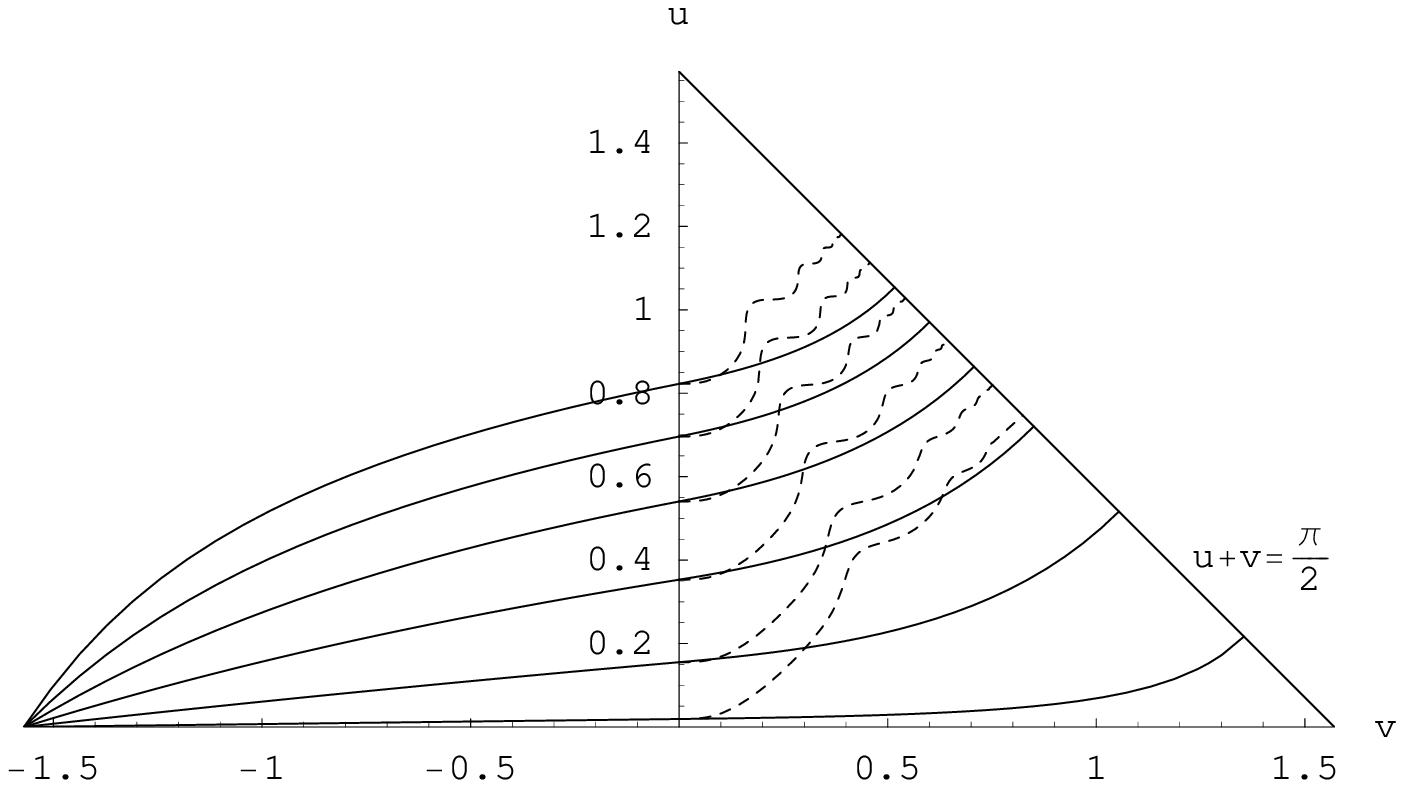}
\vspace{5mm} \caption{Figure 3
repeated with instead $K_y=3/2$. Neither the current lines nor the
geodesics change qualitatively. } \label{fig:4}
\end{figure}

In all the plots, the corresponding null geodesics are
superimposed on the current lines. These are the same in the
single wave region but evolve differently in the interaction
region. Several different mode current lines are plotted, entering
the collision wave region, i.e. a sort of mode-by-mode analytical
extension of the single wave  region current is attempted, since
the general solution  for $h_1(u,v)$ and $h_2(u,v)$ is not
available here. These current lines are also superimposed on the
geodesics. In any case, this is enough to study the general
behaviour of the process: in the interaction region (considering
the single modes for various $l$ as elementary solutions of the
Dirac equation) the single mode neutrino current is no longer
geodesic,  i.e. the helicity of the Dirac particle couples to the
spacetime curvature; for generic values of $K_x$ and $K_y$, this
current shows a typical oscillatory behaviour  that is more and
more damped as one approaches the horizon; for $K_x=0$ the
behaviour is rather different in the sense that the oscillations
are not as damped as in all the other cases, including the
companion one $K_y=0$.

It is worth noting that once again the $\partial_x$ direction has
special properties which we have studied in terms of Papapetrou
fields and Killing directions in a previous paper\cite{bcl}.

\begin{figure} [b]
\includegraphics[width=150mm,height=75mm]{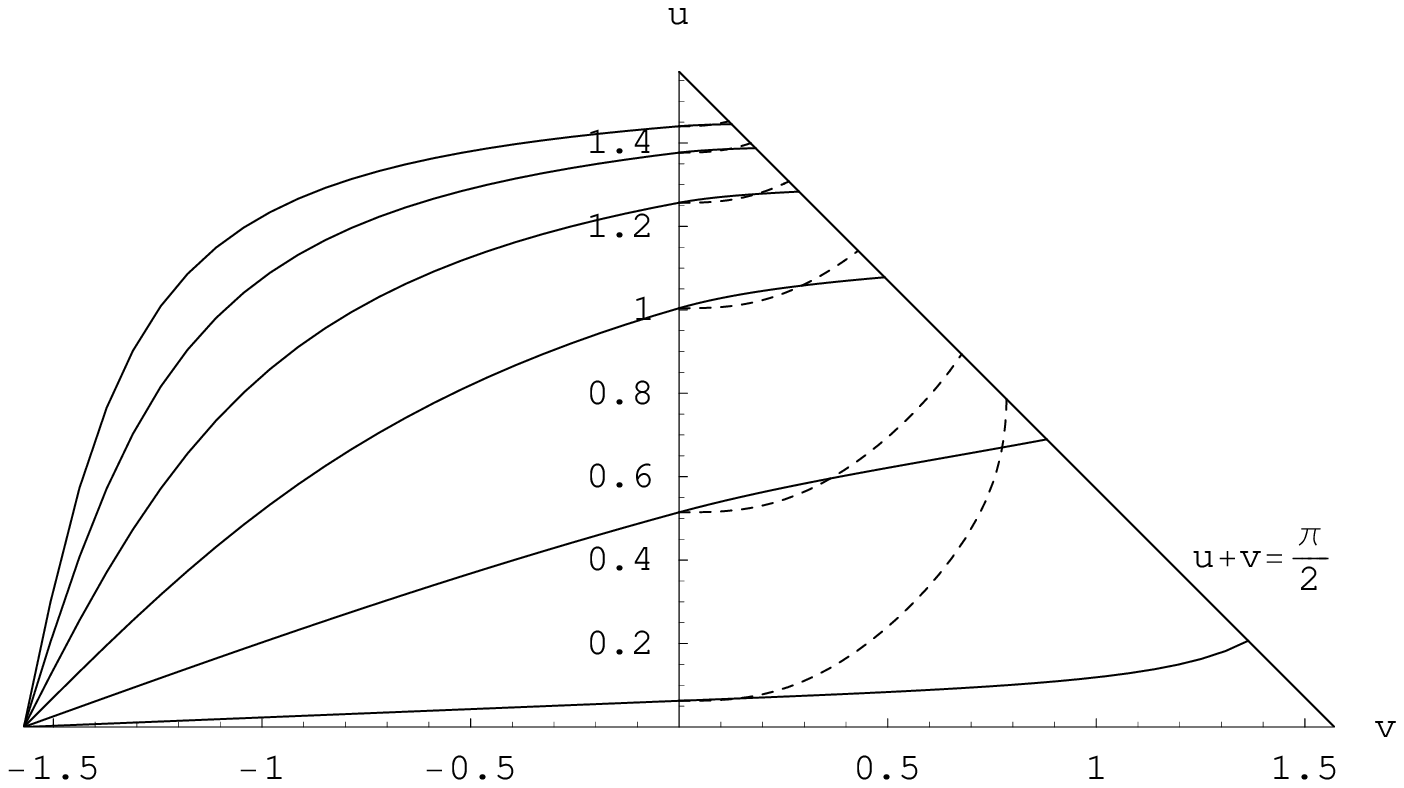}
\vspace{5mm} \caption{Figure 2
repeated for $K_x=0$, $K_y=1/2$, $0.1<|K_v|<1.1$, $l=1/2$. This
choice of $K_x$ changes the behaviour of the current field in
region I: the lines of force do not oscillate around directions
perpendicular to the horizon (i.e. around fixed values of $z$)
anymore and they are strongly different from geodesic
trajectories.} \label{fig:5}
\end{figure}

\section{Concluding remarks}

The Dirac equation in the background spacetime of two colliding
gravitational waves with parallel polarization is studied in
detail in the Newman-Penrose formalism. Explicit results have been
considered for the so called Ferrari-Iba\~nez degenerate solution.
In this case the equation is separated in all the four spacetime
regions, but the integration is complete in the single wave and
flat spacetime regions only. In the collision region instead, only
a partial analytic form of the solution can be given, involving
the spin-weighted spherical harmonics. Continuously matching the
solution across the regions' boundaries is briefly discussed too.
The characteristics of the associated neutrino current are
analysed, showing that it is geodesic in the single wave regions,
and the results are graphically summarized (see figs. 2-6).

\begin{figure} [t]
\includegraphics[width=100mm,height=75mm]{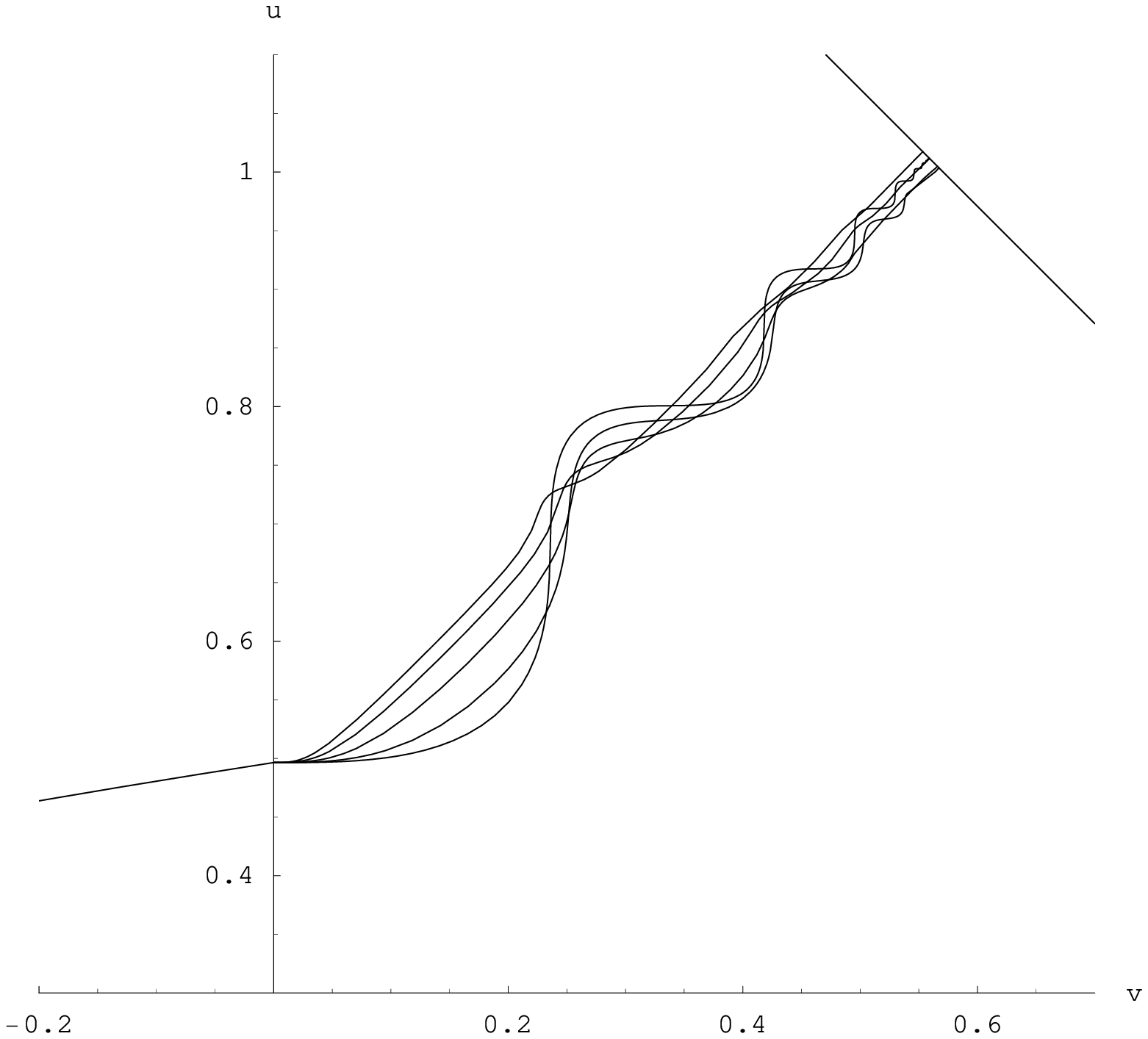}
\vspace{5mm} \caption{This
figure shows how a multiplet of current field lines forms in the
transition between region II and region I. Here $K_x=1$,
$K_y=1/2$, $|K_v|=1$ and $l$ goes from 1/2 to 9/2: the
oscillations are damped for higher values of $l$.} \label{fig:6}
\end{figure}

\begin{figure} [t]
\includegraphics[width=100mm,height=75mm]{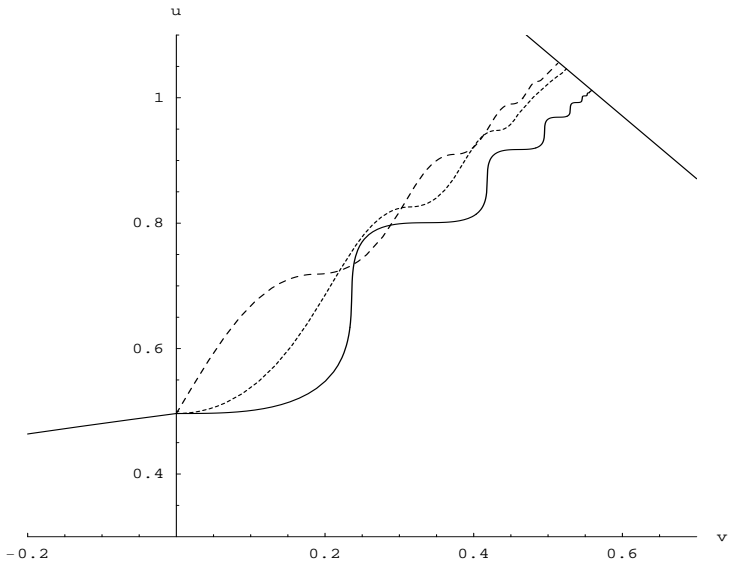}
\vspace{5mm} \caption{Here
$|h_1|$ and $|h_2|$ (the widerly and finerly dotted lines,
respectively) are plotted, with $K_x=1$, $K_y=1/2$, $|K_v|=1/8$
and $l=1/2$, and compared to the current (solid line).}
\label{fig:7}
\end{figure}

\section*{Appendix. Spin-weighted spherical harmonics}
\begin{enumerate}
\item Analytic, implementable definition
\begin{eqnarray}
 \quad {}_{s}Y^{m}_l(\theta,\,\phi)&=&
e^{i\,m\,\phi}\sqrt{\frac{2\,l+1}{4\,\pi}\;
\frac{(l+m)!\,(l-m)!}{(l+s)!\,(l-s)!}}\,\left(\sin\,\frac{\theta}{2}\right)^{2\,l} \nonumber \\
 \quad & \times
&\sum_{r=-l}^{l}(-1)^{\,l+m-r}\left(\begin{array}{c}l-s\\r-s
\end{array}\right)\left(\begin{array}{c}l+s\\r-m
\end{array}\right)\left(\cot\,\frac{\theta}{2}\right)^{2\,r-m-s}
\end{eqnarray}
\item Compatibility with spherical harmonics \begin{equation}
{}_{0}Y^{m}_l(\theta,\,\phi)=Y^{m}_l(\theta,\,\phi) \end{equation}
\item Conjugation relation \begin{equation}
{}_{s}Y^{m*}_l(\theta,\,\phi)=(-1)^{m+s}{}_{-s}Y^{-m}_l(\theta,\,\phi)
\end{equation} \item Orthonormality relation \begin{equation} \int d\Omega\;
[{}_{s}Y^{m*}_l(\theta,\,\phi)]\;[{}_{s}Y^{m'}_{l'}(\theta,\,\phi)]=
\delta_{l,l'}\, \delta_{m,m'} \end{equation} \item Completeness
relation
\begin{equation} \sum_{l,m}
\;[{}_{s}Y^{m*}_l(\theta,\,\phi)]\;[{}_{s}Y^{m}_{l}(\theta
',\,\phi ')]= \delta(\phi-\phi ')\, \delta (\theta - \theta ')
\end{equation} \item Parity relation \begin{equation}
{}_{s}Y^{m}_l(\theta,\,\phi)=(-1)^l{}_{-s}Y^{m}_l(\theta,\,\phi)
\end{equation} \item Clebsch-Gordan relation
\begin{eqnarray}
 \quad & ({}_{s_1}Y^{m_1}_{l_1})\,({}_{s_2}Y^{m_2}_{l_2})
& = \frac{\sqrt{(2l_1+1)(2l_2+1)}}
{4\pi}\sum_{l,m,s}<l_1,l_2; m_1,m_2|l_1,l_2; l,m> \nonumber \\
 \quad && \times <l_1,l_2; -s_1,-s_2|l_1,l_2; l,-s>
\sqrt{\frac{4\pi}{2l+1}}\;({}_{s}Y^{m}_{l})
\end{eqnarray}
\item Spin raising and lowering operators
\begin{eqnarray}
 \quad & {}_s\partial^+\,{}_{s}Y^{m}_l & =-\sin^s
\theta\,\left[\partial_\theta\, (\sin^{-s}
\theta\;{}_{s}Y^{m}_l)+\frac{i}{\sin \theta}\,\partial_\phi\,
(\sin^{-s} \theta\;{}_{s}Y^{m}_l)\right]\nonumber \\
  && =\sqrt{l-s}\;\sqrt{l+s+1}\;({}_{s+1}Y^{m}_l)\ ,\nonumber \\
 \quad & {}_s\partial^-\,{}_{s}Y^{m}_l &
=-\sin^{-s}\theta\,\left[\partial_\theta\, (\sin^s
\theta\;{}_{s}Y^{m}_l)-\frac{i}{\sin \theta}\,\partial_\phi\,
(\sin^s \theta\;{}_{s}Y^{m}_l)\right]\nonumber \\
  && =\sqrt{l+s}\;\sqrt{l-s+1}\;({}_{s-1}Y^{m}_l)\ .
\end{eqnarray}
\end{enumerate}
Harmonics with $s=1/2$ and different $l$ and $m$ are listed in the
following table.\\[-1cm]
\begin{center}
$$
\renewcommand{\arraystretch}{1.2}
\begin{array}{|r|l|}
\hline
(l,\,m) & {}_{1/2}Y^{m}_{l} \\[1mm] \hline
(1/2,\,-1/2) & -\frac{i}{\sqrt{2\,\pi}}\,e^{-i\,\phi/2}\,\cos(\theta/2)\\[4mm]
(1/2,\,1/2) & \frac{i}{\sqrt{2\,\pi}}\,e^{i\,\phi/2}\,\sin(\theta/2) \\[4mm]
(3/2,\,-3/2) & -\sqrt{\frac{3}{\pi}}\,i\,e^{-3\,i\,\phi/2}\,
\cos^2(\theta/2)\,\sin(\theta/2) \\[4mm]
(3/2,\,-1/2) & -\frac{i}{2\,\sqrt{\pi}}\,e^{-i\,\phi/2}\,
\cos(\theta/2)\,(3\,\cos\theta-1) \\[4mm]
(3/2,\,1/2) & \frac{i}{2\,\sqrt{\pi}}\,e^{i\,\phi/2}\,
\sin(\theta/2)\,(3\,\cos\theta+1) \\[4mm]
(3/2,\,3/2) & -\sqrt{\frac{3}{\pi}}\,i\,e^{3\,i\,\phi/2}\,
\cos(\theta/2)\,\sin^2(\theta/2) \\
\hline
\end{array}
$$
\end{center}

\end{document}